# The fusion of phonography and ideographic characters into virtual Chinese characters - Based on Chinese and English

Hongfa Zi[1], Zhen Liu[1]

Qinghai Nationalities University, School of Economics and Trade, Qinghai Province, China

The characters used in modern countries are mainly divided into ideographic characters and phonetic characters, both of which have their advantages and disadvantages. Chinese is difficult to learn and easy to master, while English is easy to learn but has a large vocabulary. There is still no language that combines the advantages of both languages and has less memory capacity, can form words, and is easy to learn. Therefore, inventing new characters that can be combined and the popularization of deep knowledge, and reduce disputes through communication. Firstly, observe the advantages and disadvantages of Chinese and English, such as their vocabulary, information content, and ease of learning in deep scientific knowledge, and create a new writing system. Then, use comparative analysis to observe the total score of the new language. Through this article, it can be concluded that the new text combines the advantages of both pictographic and alphabetical writing: new characters that can be combined into words reduces the vocabulary that needs to be learned; Special prefixes allow beginners to quickly guess the approximate category and meaning of unseen words; New characters can enable humans to quickly learn more advanced knowledge.

\* Corresponding author: 1697358179@qq.com

**Introduction**

The initial human writing was graphic writing, which is similar to pictographic writing. But in today's world, Latin alphabet is more prevalent because alphabetical writing is conducive to simple transactions, and one of the foundations of commercial prosperity in Western countries is easy to learn alphabetical writing. Alphabet writing is a commercial language that originated in the Mediterranean. Simple and easy to learn writing facilitates the exchange of commodity interests (Powell, 1989). But the short-term simplicity and ease of learning have harmed the development of some technologies. For example, the Oxford Dictionary, due to the constant emergence of new social developments and new things, has included nearly 600000 English words (Oxford Dictionary, 2010). The vast vocabulary exceeds the brain capacity of college students, unless their brain is made by a CPU. There are approximately 15000 commonly used English words (Brysbaert et al., 2016); Chinese characters are closely related to words, so the number of characters recognized can represent language ability (Zhang and Roberts, 2019), while there are only 5000 commonly used Chinese characters (Coyle, 2007). The large amount of English vocabulary can lead to many negative consequences, such as the fact that every industry is separated by a mountain, and English speakers from different industries are not familiar with each other's words (Mo, 2008); For large countries, slight differences in accents in many places can lead to alliances and social divisions (Fidrmuc and Ginsburgh, 2007), but Chinese characters are not affected by accents. But the difficulty of learning Chinese also hinders China's international exchanges, so some scholars in twentieth century want to roughly abolish Chinese characters and use letters instead.

Phonetic and ideographic characters both have representative languages, namely English and Chinese. Alphabet is a type of phonetic writing, while hieroglyphs are a type of ideographic writing, but both English and Chinese have their own advantages and disadvantages. Most Chinese characters contain prefixes (referred to as radicals by Chinese people), which can help people quickly distinguish the category of each character (Feldman and Siok, 1999), but Chinese has thousands of font patterns; Although English characters have prefixes to distinguish different parts of speech (Hammond, 1993), they do not have categories. Moreover, most words do not have prefixes, and prefixes and suffixes are not user-friendly for most beginners in English learning. Chinese in hieroglyphs has five tones in pronunciation, which can help people distinguish vocabulary and improve their ability to resist noise (Cheng, 1968). Another advantage of Chinese is its function of combination word, where each Chinese character can make up a vocabulary. Chinese word grouping equals English words (Lin and Yeh, 2015). combination word reduces the vocabulary that needs to be invented and reduces the amount of human learning. Each word in English has its own meaning, and in most cases, it is not grouped together (except for 21) (He and He, 2022). Chinese characters are monosyllabic, and their word formation function does not require the use of the connector "-". For example, when expressing the number 21, "21=二十一". The "-" symbol is not pronounced in English, and native English speakers cannot hear the connector, making it difficult to form words. Whenever a new thing appears, alphabetic writing must create a new word (Julian, 2000), and the number and load of words that people need to remember are increasing.

We need to boldly create newcharacters, combine the advantages of both, and get rid of their shortcomings. Language is the carrier of culture, and both English and Chinese have various incongruities, but their forms are almost opposite. Chinese expresses sound without meaning, while the latter expresses meaning without sound, so combining the advantages of both can overcome the shortcomings of both sides.

The newly created characters has certain reference significance for the development of both English and Chinese, and cultural exchange can also reduce many unnecessary conflicts.

|  | Stage 1 | Stage 2 | Stage 3 | Stage 4 |
|---|---|---|---|---|
| Character name | Graphic character | Wedge characters; Hieroglyphs characters; Seal characters | Alphabet characters; Hieroglyphs characters | Virtual Chinese characters |
| Character source | Natural world | Natural world | Natural world; Brain imagination | Alphabet and hieroglyphs |
| Invention time | Wilderness era | Early period of the country | Feudal modern times | Twenty-first century |
| Belonging character | Ideographic writing | Ideographic writing | Phonography writing; Ideographic writing | Expressing voice and meaning |
| User | hominid | Babylon; Ancient Egypt, Ancient China; ancient India | Greece; China | No country |

Figure 1: Characteristics of each stage of text evolution, with the main sources of text coming from nature and Brain imagination.

Text is crucial for the sustained inheritance of principles, knowledge, and rules. With the advancement of technology, people are observing more and more new things from the micro world, all things in the universe, and illusory information. It is necessary to easily learn complex knowledge with language assistance. Therefore, the creation of new characters is imperative, but this does not mean the abolition of Chinese and English. So what kind of reference material should be chosen? How to combine the advantages of English and Chinese characters? Can new writing adapt to the industrial era? How should the validity of new text be determined?

To address the many issues mentioned above, we need to first observe the advantages and disadvantages of both languages. This language system can be quickly used by both Chinese and British people and allows beginners to quickly transition to complex knowledge during learning. This way, people don't have to memorize a large number of words and can quickly guess the general category of the word; Each letter combination corresponds to a Chinese character, and after anchoring the Chinese character, the pronunciation of the letter language will not be imbalanced and will bridge the gap between Chinese and English.

This article has the following contributions: summarizing the advantages of both ideographic and phonetic writing, and creating a brand new characters based on these advantages and disadvantages. This writing system is based on the pronunciation of Chinese characters and develops new alphabetical scripts; This includes a radical table and a virtual Chinese character dictionary. The former is similar to a radical, representing the attributes of the word; The latter is the pronunciation of the font, representing the information conveyed to the audience in daily communication, and the pronunciation of the font also includes tone.

There are other parts of the paper. Firstly, there is the literature review section, which mainly presents the views of some scholars on the development of writing. Then comes the results section, which lists a list of radical pronunciations and dictionaries. The radical table is divided into radical pronunciations, whether to abolish them, radical pronunciations, and modified pronunciations; The dictionary mainly lists the specific pronunciations of 9000 Chinese characters, but due to the large number, only a partial list will be provided. Finally, put forward expectations and speculations for the future development of language.

## Literature review

### Wilderness era: The era of the graphic character

Painting was the first written language of primitive humans. Since the birth of humanity, language has existed (Shea, 2011), but the birth of systematic writing has only a history of several thousand years. Due to natural limitations, most graphics and text are relatively primitive (Harth, 1999). The source of this text is many real-life things in nature, directly imitating the specific forms of things (Grigorenko, 2023). It can directly express the meaning of graphics, similar to basic hieroglyphs, and the development of this language is also limited by human intelligence (Parker and Gibson, 1979). The writing of graphic characters is relatively difficult because it requires a certain cognitive and imaginative talent, and most of the written symbols are preserved in stone carving form (Tyl é n et al., 2020). It is not easy to express abstract things, such as time, emotions, etc., and cannot well preserve information from history and summarize lessons learned (Coolidge and Wynn, 2018). Primitive humans used biological instincts to acquire knowledge, which resulted in their weak ability to organize knowledge (Vahia, 2016), partly because humans did not have suitable written carriers in the primitive era, and knowledge was inevitably distorted in the process of inheritance. The tools and systems of primitive humans were not well-developed, so when planning the future using symbols and writing, it was necessary to use stones for carving and recording (Henshilwood et al., 2002). All of these situations indicate that primitive humans continued to improve characters.

**Early period of the country: cuneiform and hieroglyphs period**

Cuneiform and hieroglyphs were one of the earliest human characters. The cuneiform script originated in the Mesopotamian plain and was founded by the Sumerians (Powell, 1981). With the development of this type of writing, not only Sumerians, but also Persians, Akkhs, and Hittites used cuneiform script (Cohen, 2006). The language gradually became outdated during its development and evolved into another form - alphabetic characters (Ullman, 1927). But its cousin hieroglyphs continue to evolve in China and Egypt. However, due to the emergence of cultural chronology in Egypt, the interpretation of its hieroglyphs became obscure and difficult to understand (Fischer, 1977); The ancient Indian script was a mixed font of hieroglyphs and alphabets, and it has also been lost. The initial hieroglyphs originated from various tools and human body movements (Jie and Xiaoming, 2014). The primitive hieroglyphs in China evolved from inscriptions and patterns on natural objects and pottery in primitive societies (Lu and Aiken, 2004). Later Chinese characters were based on the first generation of hieroglyphs and created new characters using methods such as understanding, pointing, and sound (DeFrancis, 1986), and the radical parts of Chinese characters were gradually used. For the sake of governance, ancient Chinese rulers preferred a writing system that was not influenced by local accents and was easy to pass the imperial examinations. Regardless of how dialects in various parts of China changed, knowledge of hieroglyphs could be continuously accumulated and interpreted (Chang, 1989). A typical feature is that Old English 1500 years ago was difficult for modern British people to interpret because the letters and grammar of both languages had changed (Freeborn, 1998); However, ancient Chinese can be interpreted by modern Chinese people (Needham et al., 1975) because Chinese has natural semantics and it is difficult to change the shape of characters.

**Feudal modern times: The period of alphabets and hieroglyphs**

In feudal modern times, the great development of alphabets resulted in only Chinese being left as a pictographic language. Alphabet characters has spread throughout the world with the development of Western civilization. North America, South America, Africa, and Oceania are all countries that use alphabets, leaving only China to still use hieroglyphs (Liang B et al., 2010). In the early stages of learning, alphabetical writing can quickly learn and invent words, which is beneficial for encouraging more people to participate in commodity trading (Grossmiller, 2004). Alphabet learning requires the use of images to recognize words (Piro, 2002), while hieroglyphs are not very necessary because graphics have certain meanings. China is very backward in modern times, and some people attribute part of the responsibility to a low literacy rate (Wan, 2014). Later, China promoted the simplification of characters and the pinyin recognition movement, which divided letters into five tones. People could search for letters in dictionaries and correct pronunciation, achieving the popularization of Mandarin (Zhao and Baldauf, 2011). In addition, Chinese characters have different vowels, similar to "aeiou", where the pronunciation of the vowels changes from one letter to two letters, such as "en" (Hua and Dodd, 2000). After the reform and opening up, Chinese transliterated a large number of foreign words from English according to semantics, such as poker and shafa, and gradually used 26 fonts of English as a foreign language (HU and PETER, 2004). However, due to the fact that Chinese characters are not suitable for the development of the electronic age: $2^8-1=255$ font shapes, there are several thousand commonly used Chinese characters ($2^8=256$ font shapes, cannot represent 8 zeros or 1s), and Chinese characters also have limitations.

In summary, each scholar has different perspectives on exploring the development of writing. The development of national characters has shown three directions, namely phoneticcharacters, ideographic characters, and a mixture of the two. However, each hieroglyph cannot accurately correspond to a letter word, and the combination of letter words is always a big problem. Therefore, some humble suggestions should be provided for the development of modern writing, and a reference point should be provided for the future development of linguistics.

This article attempts to design a method that combines the advantages of two languages, which has some impact on the development and evolution of future languages. The key to exploring in the article is the pronunciation modification of related radicals, which is crucial for solving repetitive Chinese character pronunciation letters, and considering the benefits of new virtual Chinese characters for humans, such as whether they can help learning and create more scientists.

## Results

The new character is superior to the original two in terms of information content, accuracy, and scientific knowledge learning. However, when grouping words, the characters occupied by the new text are slightly longer than those in English (words can be omitted and vowels can be reduced when used). The radical letters can accurately distinguish the category of each word, and their learning difficulty is lower than that of Chinese characters, which is conducive to more people learning to speak Chinese in a short period of time.

### Initial overall plan

According to the Chinese characters in the Xinhua Dictionary, corresponding letter words are formulated, characterized by each Chinese character having a corresponding letter combination and Each letter combination can make up a new vocabulary. Each character (including "·, ´, -") is a character on the Ascii character table and its extension table, which can be represented by 8 bits (8 0 or 1, for example, 01000001 corresponds to A)

(1) Virtual Chinese characters are composed of radical phonetic letters and the letters that represent the pronunciation of the Chinese characters themselves. The representative letters of the radical and the pronunciation of the Chinese characters are connected by the "·" symbol, forming a single virtual Chinese character, for example, "协" corresponds to "shi·xié". This can distinguish categories through radicals; (2) Individual virtual Chinese characters are connected to each other through a "-" symbol to form new words; (3) The radical letters and the letters that sound the Chinese characters themselves have five tones, representing the light tone, first tone, second tone, third tone, and fourth tone, respectively. For example, "Ee Ēē Éé Ěě Èè " Change to "Ee Ëë Éé Êê Èè" (tone can reduce repeated letters, and fewer syllables can express more words); ǘ Change to "uú" mainly because the pronunciation of the radical of "女" is "nǚ" and "ūūúǔǜ" is not on the ASCII character table; (4)If there is still repetition after using five tones, it can be prevented by using the letters+'+the pronunciation after removing the radical+the pronunciation letter of the Chinese character itself, such as 䑌 "yue'juan·téng". However, in a dictionary of nearly 10000 words, there are only a few rare characters; (5) To avoid repetition of radical pronunciations, the pronunciations of radicals were modified, such as "石, 尸", changing their pronunciations as radicals to "bo, po". The main function of radical letters is to distinguish categories; (6) In order to be easy to learn and remember, letters with similar radicals and initial letters are used, such as "石, 尸" respectively corresponds to "po, bo" ; (7) Part of the radicals only have a few words, such as "己, 已, 巛" and so on. Abolish them to prevent the repetition of radical pronunciations,。such as the pronunciation of "衣" and "已" is repetitive, with the latter being abolished; (8) The upper and left radicals are preferred for radicals, except for radicals such as "鸟, 鱼, 犭", such as 鹰, where 鸟 are not chose 广; (9) The numbers "一、二 ,三、四、五、六、七、八、九、十、百、千、万、亿、兆、京" are composed of their own pronunciation letters and have tones; (10) Abbreviate the formed letters, such as "ing, ong" to "ig, og"; Reduce "uang" to "ug". (11) Radical letters can be pronounced or not pronounced; The "-" and "·" symbols can also be pronounced or not pronounced. Because the radicals of Chinese characters are also not pronounced when speaking, but Chinese people can also understand them.

**Sub scheme: Chinese radical table**

The Chinese radical table is divided into original radical letters and modified radical letters, with the latter being more important. The original radical comes from the Xinhua Dictionary, and the pronunciation of radical has a large number of identical phonetic letters (phonetic letters in silent tones), so it needs to be modified. The modified radical has different pronunciations, but synonymous radicals can use the same pronunciation, such as "yan" for both "言" and "讠".

Part of the inspiration comes from the Xinhua Dictionary. (1)The numbers in the first column: The numbers in the first column in the following figure are the same as those in the Xinhua Dictionary. You can search for radicals based on the number in the first column, but due to the hundreds of pronunciations of radicals, only a portion is displayed. (2)Color difference: Consistent pronunciation and synonymous radicals have the same color, which can maintain differentiation. (3)Abolish partial radicals: Due to the fact that some radicals only have a few words, it is not worth setting up a radical. After abolishing radicals, the corresponding characters are classified as other radicals.(4) Revised Pronunciation: Due to the presence of too many repeated homophonic radicals, some radicals cannot be their own pronunciation. For example, when "尸" reads "shi", it is changed to "po". (5)Extra words: For the convenience of adjectives and adverbs, the original pronunciations of "地" and "的" are retained in the following figure.

Radical table (partial, all at the bottom)

| The radical of the Xinhua Dictionary | Pronunciation of the radical itself | After abolition, it will be classified as other radicals | Revised Pronunciation (Important) | Number of repetitions |
|---|---|---|---|---|
| One stroke radical | | | | |
| 1一(héng) | heng | | h | 1 |
| | | | heng | 1 |
| 2丨(shù) | shu | | s | 1 |
| | | | shu | 1 |
| 3丿(piě) | pie | | p | 1 |
| | | | pie | 1 |
| 4丶(diǎn) | dian | | d | 1 |
| | | | dian | 1 |
| 5乛(zhé)乙⺃乚 | zhe | | z | 1 |
| | | | zhe | 1 |
| Two stroke radical | | | | |
| 6十(shí) | shi | | shi | 1 |
| ナ(zuǒ) | zuo | | wan | 1 |
| 29(士)(shì) | shi | | do | 1 |
| 51尸(shī) | shi | | po | 1 |
| 102石(shí) | shi | | bo | 1 |
| 110矢(shǐ) | shi | | ro | 1 |
| 89氏(shì) | ti | Abolish (with fewer fonts) | | 0 |
| 100示(hì) | i | | i | 2 |
| 100(礻)(shì) | i | | i | 2 |
| 7厂(chǎng) | an | | an | 2 |
| 7(厂)(fǎn,yì) | an | | an | 2 |
| 46广(guǎng) | en | | en | 1 |
| Extra words | | | | 0 |
| 的de | 的 | | de | 0 |
| 地dì | 地 | | dì | 0 |

Figure 2: The radical pronunciation in the first column comes from dictionaries and online searches; The pronunciation of the remaining rows and columns comes from creation. Pink represents repetition

**Sub scheme: Chinese Dictionary Table**

The tone of alphabets objectively exists. Due to the limited length of this article, 22 Chinese characters were selected as examples, and they have tones. For example, English does not have a tone when writing, but has a tone when communicating, so phonetic symbols are also needed to assist. New writing does not require phonetic symbols when it has a tone.

Words with the same radical have the same pronunciation and must retain their tone. For example "Ee Ëë Éé Êê Èè". This method can solve a large number of Chinese characters with the same pronunciation and radical; such as "肢", "脂", and "胝", the radical pronunciation of the three characters is changed to "yue·zhǐ" (originally yue·zhī); "yuë·zhǐ" (originally yuë·zhī), "yué·zhǐ" (originally "yué·zhī). This way, there is no need to modify the pronunciation of the Chinese characters themselves: the tones are arranged according to the number of uses, and "肢" are used the most, so they are ranked first (light tone), "脂" is ranked second (first tone), and "胝" is ranked third (second tone). The pronunciations of "人" and "亻" are both "ren", but there are too many characters with the radical of "人", so people directly use the three tone "rén", such as "合" corresponds to "rén·hé" ,not "ren·hé".

The table of Chinese characters    (partial, all at the bottom)

| Chinese characters and pronunciations | Chinese characters | Pronunciation with radical (red represents repetition) | Number of repetitions | Characters after tone modification (can be represented by 8 bits) | After removing n from "ing", "ong", etc (Final version) |
|---|---|---|---|---|---|
| 讠 yán | 讠 | yan |  | yan | yan |
| 计yan·jì | 计 | yan·jì | 2 | yan·jì | yan·jì |
| 订yan·dìng | 订 | yan·dìng | 1 | yan·dìng | yan·dìg |
| 记yan·jì | 记 | yan·jì | 2 | yän·jì | yän·jì |
|  |  |  |  |  |  |
| 一héng |  | h | 1 | h | h |
| 一yī | 一 | yī | 1 | yī | yī |
| 二èr | 二 | èr | 1 | èr | èr |
| 丁h·dīng | 丁 | h·dīng | 1 | h·dīng | h·dīg |
| 丽h·lì | 丽 | h·lì | 2 | h·lì | h·lì |
| 吏h·lì | 吏 | h·lì | 2 | hëng·lì | hëg·lì |
|  |  |  |  |  |  |
| 十shí |  | shi | 1 | shi | shi |
| 十shí | 十 | shí | 1 | shí | shí |
| 支shi·zhī | 支 | shi·zhī |  | shi·zhī | shi·zhī |
| 协shi·xié | 协 | shi·xié | 1 | shi·xié | shi·xié |
|  |  |  |  |  |  |
| 士shì |  | shi | 2 | do | do |
| 士do·shì | 士 | do·shì | 2 | do·shì | do·shì |
| 吉do·jí | 吉 | do·jí | 1 | do·jí | do·jí |
| 壳do·ké | 壳 | do·ké | 1 | do·ké | do·ké |
|  |  |  |  |  |  |
| 尸shī |  | shi | 1 | po | po |
| 尸po·shī | 尸 | po·shī | 1 | po·shī | po·shī |
| 屡po·lǚ | 屡 | po·lǚ | 2 | pö·luǔ | pö·luǔ |
| 履po·lǚ | 履 | po·lǚ | 2 | po·luǔ | po·luǔ |

Figure 3: The radical pronunciation in the first column comes from dictionaries and online searches; The pronunciation of the remaining rows and columns comes from creation. We only need to learn 5000 characters (final version) to handle most of the knowledge, even if you go from undergraduate to graduate.

**Sub scheme: The table of combination vocabulary**

Letters combination can be form words. Individual virtual Chinese characters are connected to each other through the "-" symbol to form words, which can greatly reduce the number of vocabulary that needs to be learned. Ordinary college students only need to learn a few thousand words and letter combinations to quickly achieve the learning of scientific knowledge and the construction of knowledge systems. The most important thing is to achieve interdisciplinary communication . This approach can cultivate more talents

The table of combination vocabulary

| Chinese vocabulary | English vocabulary | Corresponding letters | Revised letter (Final version) |
|---|---|---|---|
| 橡皮 | rubber | mu·xiàng pi·pí | mu·xiàg-pi·pí |

| 为了 | in order to | d·wéi-z·le | d·wéi-z·le |
|---|---|---|---|
| 狐狸 | fox | a·hú-a·lí | a·hú-a·lí |
| 再见 | goodbye | h·zài-ti·jiàn | h·zài-ti·jiàn |
| 高兴 | happy | gao·gāo-bǎ·xīng | gao·gāo-bǎ·xīg |
| 礼帽 | formal hat | i·lǐ-nin·mào | i·lǐ-nin·mào |
| 喂 | hello | kou·wèi | kou·wèi |
| 嗨 | hey | kou·hāi | kou·hāi |
| 多少 | how much | zo·duō-ǎo·shǎo | zo·duō-ǎo·shǎo |
| 几岁 | How old are you | ji·jǐ-sa·suì | ji·jǐ-sa·suì |
| 布娃娃 | cloth dolls | wan·bù-nü·wá-nü·wá | wan·bù-nü·wá-nü·wá |

Figure 6: Each Chinese character vocabulary in the table corresponds to a letter combination. This can change Chinese to Latin letters without changing its pronunciation.

We can further add more new vowel letters, such as changing 'ing' to 'ŋ' and setting 'ŋ' as a new vowel; Change abbreviations such as' uang 'to' ug '; Computers can be used to assign the partial pronunciations of fewer letters to more characters (under the premise of easy learning of pseudonyms), so that the text has more variations. As shown in the table below:

Translation table for three languages

| 在信息产业和复杂制造中，由于仪器的多而复杂、技术人员的精准分工，企业新技术的诞生和价值增长更加依赖技术工人的自我驱动和技术同事的监督检验。 |
|---|
| In the information industry and complex manufacturing, due to the abundance and complexity of instruments and the precise division of labor among technical personnel, the birth and value growth of new technologies in enterprises rely more on the self drive of technical workers and the supervision and inspection of technical colleagues. |
| wan·zài ren·xìn-xīn xī le·chân-re·yè hë·hé zhi·fù-pin·zá dao·zhì-zou·zào s·zhöng, s·yóu-h·yú ren·yí-kou·qì de zo·duö h·ér zhi·fù-pin·zá, ou·jì-mu·shù rén·rén-kou·yuán de mi·jīng-han·zhūn ba·fēn-gi·gōng, rén·qî-re·yè qin·xīn ou·jì-mu·shù de yan·dàn-sen·shēng hë·hé ren·jià-ren·zhí tu·zēng-h·zhâng h·gèng-lì·jiä ren·yī-dan lài ou·jì-mu·shù gi·gōng-rén·rén de p·zì ge·wô ma·qü-lì·dòng hë·hé ou·jì-mu·shù to·tóng-hëng·shì de min·jiän-mo·dü mu·jiân-ma·yàn. |
| This is a translation that has been compressed and allocated by the computer |
| shî·zài re·xìn-xin xĭ le·chân-en·yè hë·hé zui·fù-pin·zá dao·zhì-ou·zào s·zhög , s·yóu-h·yú re·í-o·qì de xa·dö h·ér zui·fù-pin·zá , e·jì-i·shù rë·rén-o·yuán de mi·jīg-han·zhūn bâ·fēn-no·gög , rê·qî-en·yè jun·xīn e·jì-i·shù de yan·dàn-sen·shēg hë·hé re·jà-re·zhí tu·zēg-h·zhâg h·gèg-li·jä re·ï-dan lài e·jì-i·shù no·gög -rê·rén de p·zì te·ô ma·qü-li·dòg hë·hé e·jì-i·shù to·tóg-hëg·shì de min·jän-mu·dü i·jân-ma·yàn. |

Figure 6: Mutual translation of three languages, namely English, Chinese, and Virtual language. The virtual language converter is located in another data file. It is similar to a translator that can transform input Chinese into virtual language. Its source code is saved in the uploaded notepad.

**Simple comparison table**

Characters needs to be compared to determine its superiority or inferiority. There are three levels available for comparison and ranking, calculating the total score, and allowing for simultaneous ranking of three points. The reason is that virtual Chinese characters have some common features at certain times, so they should have the same ranking. For example, the level of ease of learning, both belong to alphabetical writing, so both are scored 3 points.

It is expected that the language will take four months to learn and complete 4000 virtual Chinese characters, completing daily communication.

|  | Shallow knowledge Learning | Shallow vocabulary | information content | Easy to learn level | Deep knowledge learning | Deep vocabulary | Whether to combine vocabulary | Do you need to create new characters | Is it beneficial for trade |
|---|---|---|---|---|---|---|---|---|---|
| English | 3 | 3 | 1 | 3 | 1 | 3 | 1 | 1 | 3 |
| Chinese | 1 | 1 | 3 | 1 | 2 | 1 | 3 | 3 | 1 |
| Virtual Chinese characters | 3 | 2 | 2 | 3 | 3 | 2 | 2 | 3 | 2 |

English: 3+3+1+3+1+3+1+1+3=19

Chinese: 1+1+3+1+2+1+3+3+1=16

Virtual Chinese characters: 3+2+2+3+3+2+2+3+2=22

From the above observations, virtual Chinese characters have the highest score.

### Discussions

Both alphabets and hieroglyphs have not undergone revolutionary changes since modern feudal times. Although there are only 5000 commonly used characters and patterns in Chinese, there are a total of 20000 characters and patterns in the dictionary. In modern times, China wanted to abolish simplified characters and switch to Latin letters (Pang and Ko, 2019), but after a series of activities to improve literacy rates, the idea changed (Joan, 1992). But Chinese is still one of the most difficult languages to learn globally, as characters patterns of thousands are different, and it has hindered China's financial and technological development (Huang and Ma, 2007). The decentralized character creation model in English allows inventors of things to create new characters (Algeo, 1980). In June 2009, a study showed that the number of English vocabulary had reached one million (Knowles, 2010), partly because English absorbed many words from foreign languages that British people are not familiar with it . Therefore, it is necessary to reduce some words and unify the management of word creation rights. Unified and easy to learn characters can help people learn more knowledge. Compared to previous articles, this article focuses more on the integration of Chinese and English, such as converting Chinese Pinyin into one-to-one corresponding letter combinations; Advocate the development of radical pronunciations for English letters and increase their word formation function. English and Chinese should not be hostile, but should learn from each other in the process of development.

### Method

Literature research method: The main research direction of this article is in the field of characters, so references were made to the Oxford Dictionary and the Xinhua Dictionary. The text and vocabulary in these two books are very helpful for studying new characters. And use about 40 historical documents found on Google Scholar to search for the history of text development and the opinions of Chinese and Western scholars, evaluate the advantages and disadvantages of major texts in history, and refer to the 9000 Chinese characters in the Xinhua Dictionary to create corresponding virtual Chinese characters word by word, providing preparation for the overall design and comparative research in the future.

Comparative analysis method: Evaluate the quality of a new language from 8 indicators. They are shallow knowledge learning, shallow vocabulary, information content, ease of learning, deep knowledge learning, deep vocabulary, whether to form words, whether to create new words, and whether it is beneficial for trade. Analyze the relationship between the two parties and divide it into three levels, calculate the total score, and evaluate the quality of the language in a simple way.

### Future predictions and expectations

Language and writing, as tools for mutual communication in human society, carry the development of human society. In the future, language will further integrate, and its corresponding shortcomings will become fewer and fewer. I hope to add more new vowel letters in the near future, such as changing ing to ŋ, take ŋ Establish as a new vowel; The consonant vowel combination table is an alphabet invented by the Chinese people. "a, e, i, o, u, ing,,iong,uang, an, ang, and so on all" have the same function, similar to the vowel letters in English, with a total of 24; We can change abbreviations such as "uang" to "ug", so that the characters has more variations; Computers can be used to assign the partial pronunciations of fewer letters to more characters (on the premise of easy learning of virtual characters), so that there are more variations in the text.

### Data sources

The data is sourced from Xinhua Dictionary, Oxford Dictionary, Chinese consonant and vowel combination table, Ascii character table, and Ascii extended table. The Xinhua Dictionary originated from the 12th edition of the Xinhua Dictionary and was published by the Commercial Press; The source words of the Oxford Dictionary are from the 10th edition of the Oxford Advanced English Chinese Dictionary, as well as the Ascii character table and extension table https://www.asciim.cn/ ; The data is sourced from Xinhua Dictionary, Oxford Dictionary, Chinese consonant and vowel combination table, Ascii character table, and Ascii extended table. The Xinhua Dictionary originated from the 12th edition of the Xinhua Dictionary and was published by the Commercial Press; The source words of the Oxford Dictionary are from the 10th edition of the Oxford Advanced English Chinese Dictionary, as well as the Ascii character table and extension table (https://www.asciim.cn/) ; The consonant vowel combination table is derived from the Chinese elementary school pinyin table.